\documentstyle[12pt]{article}

\newcommand{\beq}{\begin{equation}}
\newcommand{\eeq}{\end{equation}}
\newcommand{\beqn}{\begin{eqnarray}}
\newcommand{\eeqn}{\end{eqnarray}}
\newcommand{\slp}{\raise.15ex\hbox{$/$}\kern-.57em\hbox{$\partial
$}}
\newcommand{\lnA}{\raise.15ex\hbox{$/$}\kern-.57em\hbox{$A$}}
\newcommand{\lnB}{\raise.15ex\hbox{$/$}\kern-.57em\hbox{$B$}}
\newcommand{\bP}{\bar{\Psi}}
\newcommand{\slU}{\raise.15ex\hbox{$/$}\kern-.57em\hbox{$U$}}

\newcommand{\hs}{\hspace*{0.6cm}}

\begin{document}
\title{ Field theory approach to one-dimensional electronic systems \thanks{Talk delivered
at {\it Trends in Theoretical Physics - CERN - Santiago de Compostela -
La Plata Meeting, La Plata, April-May 1997}, to be published}}
\author{ Carlos M. Na\'on
\thanks{Investigador CONICET, Argentina}
\\
{\normalsize\it
Departamento de F\'\i sica, Universidad Nacional de La Plata}\\
{\normalsize\it
C.C. 67, 1900 La Plata, Argentina}}
\date{}
\maketitle

\begin{abstract}
In this talk I describe a field-theoretical approach
recently introduced in refs.\cite{1}-\cite{3}, that can be 
used as an alternative framework to study one-dimensional systems of 
highly correlated particles. 
\end{abstract}
\newpage
\tableofcontents
\newpage

\section{Motivations}

In this talk I describe recent work on the bosonization of a non-local
Quantum Field Theory (QFT) and its application to the study of one-dimensional
(1d) many-body systems. This research has been done in collaboration with 
{\bf Virginia Man\'{\i}as}, {\bf Marta Trobo} and {\bf Cecilia von Reichenbach} from the University of 
La Plata, {\bf Daniel Barci}  from Universidade do Estado do Rio de Janeiro 
(UERJ), Brasil, and {\bf Kang Li} from Hangzhou University, China.

{}~
Since Mattis and Lieb \cite{Mattis} showed how to correctly quantize the
model of 1d fermions proposed by Luttinger \cite{Luttinger} (which was nothing 
but a slightly modified version of the model introduced by Tomonaga many years before \cite{Tomonaga}), the study of the highly-correlated electronic liquid remained an outstanding problem in mathematical physics. Perhaps the main reason for this sustained interest has been the rather strange behavior of the 2-point fermionic correlator which,
in 1d presents a branch cut instead of a pole structure. Consequently one cannot define one single particle states , meaning that the usual quasiparticle (Landau) picture fails. One then expects that only collective modes (CDW4s and SDW4s) will be present in the spectrum. These features, together with the separation of spin and density waves and the disappearence of the Fermi edge in the momentum distribution, characterize the so called Luttinger liquid behavior \cite{Haldane}. During the last decade a more practical and surprising motivation has been added to the previous academic
reasons to investigate the 1d electronic system. Indeed, striking developments in the field of nanofabrication have allowed to build real 1d semiconductors \cite{nano} (Please see \cite{Voit} for a review of recent theoretical developments). Stimulated by this curious situation in which simplified and unrealistic models became closely related to reality, we tried to employ some path-integral techniques, which have been very useful in the study of $1+1$ QFT4s, in order to provide an alternative field-theoretical approach to the study of 1d many-body systems. This led us to propose a non-local generalization of the Thirring model which contains the Tomonaga-Luttinger model as a particular case. In the next section we shall show how to deal with such a non-local theory.
\newpage

\section{The Non-Local Thirring model}

\hs We start by considering the Euclidean vacuum functional

\beq
Z = N \int D\bP D\Psi~ e^{-S}
\label{1}
\eeq

\noindent where N is a normalization constant and S is given by

\beq
 S = \int d^2x~ \bP i \slp \Psi  - \frac{g^2}{2} \int d^2x d^2y ~
[V_{(0)}(x,y) J_0(x) J_0(y) + V_{(1)}(x,y) J_1(x) J_1 (y)]
\label{2}
\eeq

\noindent where $J_{\mu} = \bP \gamma_{\mu} \Psi$ and 
$ V_{(\mu)}(x,y)$
is an arbitrary function of two variables. Note that for 
$V_{(0)} = V_{(1)} = \delta^2(x-y)$ one recovers the usual covariant Thirring
model.

As it is habitual in the path-integral approach to the
usual Thirring model, one eliminates this quartic fermionic interaction
by introducing an auxiliary vector field $A_{\mu}$. As we shall see, 
in the present non-local case one needs one more auxiliary field to achieve
the same goal. In order to depict this procedure we first observe that
S can be splitted in the form

\beq
S = S_0 + S_{int}
\label{3}
\eeq

\noindent where

\beq     
S_0 = \int d^2x~ \bP i\slp\Psi,
\label{4}
\eeq

\noindent and

\beq
S_{int} = -\frac{g^2}{2} \int d^2x~ J_{\mu} K_{\mu}.
\label{5}
\eeq

\noindent In this last expression $J_{\mu}$ is the usual fermionic
current,

\beq
J_{\mu}(x) = \bP(x)\gamma_{\mu} \Psi(x),
\label{6}
\eeq

\noindent and $K_{\mu}$ is a new current defined as

\beq
K_{\mu}(x) = \int d^2y~ V_{(\mu)}(x,y)J_{\mu}(y).
\label{7}
\eeq

\noindent Please note that no sum over repeated indices is implied
when a subindex $(\mu)$ is involved. Introducing a functional delta the partition function can be expressed as

\beq
Z = N \int D\bP D\Psi D\tilde{A}_{\mu}D\tilde{B}_{\mu}~ exp[-\{S_0 + 
\int d^2x [\tilde{A}_{\mu}\tilde{B}_{\mu} - \frac{g}{\sqrt{2}}(\tilde{A}_{\mu}
J_{\mu} + \tilde{B}_{\mu}K_{\mu})] \}]
\label{8}
\eeq

\noindent On the other hand, the fermionic piece
of the action can be written as

\beq
S_0 - \frac{g}{\sqrt{2}}\int d^2x~(\tilde{A}_{\mu}J_{\mu} + \tilde{B}_{\mu}
K_{\mu}) =
\int d^2x~ \bP [i\slp - \frac{g}{\sqrt{2}}\gamma_{\mu}
(\tilde{A}_{\mu} + \bar{B}_{\mu})] \Psi,
\label{10}
\eeq

\noindent where we have defined 
\beq
\bar{B}_{\mu}(x) = \int d^2y~ V_{(\mu)}(y,x)\tilde{B}_{\mu}(y).
\label{11}
\eeq

\noindent For later convenience we shall invert (\ref{11}) in the form

\beq
\tilde{B}_{\mu}(x) = \int d^2y~ b_{(\mu)}(y,x) \bar{B}_{\mu}(y),
\label{12}
\eeq

\noindent with $b_{(\mu)}(y,x)$ satisfying  

\beq
\int d^2y~ b_{(\mu)}(y,x) V_{(\mu)}(z,y) = \delta^2 (x-z).
\label{13}
\eeq

\noindent Equation (\ref{10}) suggests the following change of auxiliary
variables:

\beq
\frac{1}{\sqrt{2}}(\tilde{A}_{\mu} +\bar{B}_{\mu}) = A_{\mu},
\label{15a}
\eeq       
    
\beq
\frac{1}{\sqrt{2}}(\tilde{A}_{\mu} - \bar{B}_{\mu}) = B_{\mu}.
\label{15}
\eeq

\noindent From now on we shall restrict our study to the case in which 
the bilocal
functions $V_{(\mu)}$ and $b_{(\mu)}$ are symmetric. Under these conditions the partition
function of the system is given by

\beq
Z = N_1 \int DA_{\mu} DB_{\mu}~ det(i \slp + g \lnA) e^{-S[A,B]},
\label{17}
\eeq

\noindent where $S[A,B]$ is such that A and B are decoupled. Moreover, B is 
not coupled to fermionic fields either, and can then be absorbed in the 
normalization constant. (Besides, B has negative metric and must be 
eliminated in order to have a good Hilbert space. The appearence of this 
ghost is not due to non-locality, it is already present in the local model \cite{Klaiber}).
Thus we have been able to express $Z$ 
in terms of a fermionic determinant:
\beq
Z = N_2 \int DA_{\mu} det(i \slp + g \lnA) e^{-S[A]},
\label{18}
\eeq

\noindent where $N_2$ includes the contribution of the "non-local ghost" 
$B_{\mu}$ 
and $S[A]$ will be rewritten in terms of two new scalars $\Phi$ and $\eta$ by using
\beq
A_{\mu}(x) = \epsilon_{\mu\nu}\partial_{\nu}\Phi(x) + \partial_{\mu}\eta(x).
\label{21}
\eeq

\noindent At this stage one can employ the machinery of the path-integral approach to bosonization, based on a chiral change in the fermionic
path-integral measure with $\Phi$ and $\eta$ as parameters. Taking into account the corresponding Jacobian we finally get

 \beq
 Z = N \int D\Phi D\eta~ e^{-S_{eff}}
 \label{27}
 \eeq

 \noindent where 

 \beqn
 S_{eff}&=&\frac{g^2}{2\pi} 
                        \int d^2x~ (\partial_{\mu}\Phi)^2 +\nonumber\\
        &+& \frac
        {1}{2}\int d^2x d^2y [b_{(0)}(y,x) \partial_1 \Phi (x)
        \partial_1 \Phi (y) + b_{(1)}(y,x) \partial_0\Phi (x)
        \partial_0 \Phi (y)] + \nonumber\\
        &+&
        \frac{1}{2}\int d^2x d^2y [b_{(0)}(y,x) \partial_0 \eta (x)
        \partial_0 \eta (y) + b_{(1)}(y,x) \partial_1 \eta (x)
        \partial_1 \eta (y)] + \nonumber\\
        &+&
        \int d^2x d^2y [b_{(0)}(y,x) \partial_0 \eta (x)
        \partial_1 \Phi (y) - b_{(1)}(y,x) \partial_1 \eta (x)
        \partial_0 \Phi (y)]
 \label{28}
 \eeqn

\noindent Equations (\ref{27}) and (\ref{28}) constitute the main 
result of this Section.
Thus, we have been able to extend the path-integral approach to bosonization, 
 previously applied to the solution of local QFT's, to a Thirring-like model
 of fermions with a non-local interaction term. More specifically, we 
 have shown the equivalence between the fermionic partition function
(\ref{1}) and the functional integral (\ref{27}) corresponding to the two 
bosonic degrees of freedom $\Phi$ and $\eta$ with dynamics governed by
(\ref{28}). The contribution to this action coming from the fermionic 
Jacobian (the first term in the r.h.s of (\ref{28})) exactly coincides
with the one which is obtained in the local case. On the other 
hand, the effect of non-locality is contained in the remaining terms,
through the inverse potentials $b_{\mu}(x,y)$. Note that, even in the non-local 
case, $\Phi$ and $\eta$ become decoupled for $b_{(0)} = b_{(1)}$. Of course,
when $b_{(0)} = b_{(1)} = \delta^2(x - y)$, one recovers the bosonic version
of the local Thirring model.

 The spectrum of this bosonic model can be more easily analyzed in momentum
 space. Indeed, by Fourier transforming (\ref{28}) one obtains

 \beqn
 S_{eff}& =& \frac{1}{(2\pi)^2} \int d^2p \{\hat{\Phi}(p) \hat{\Phi}(-p)
           A(p) \nonumber \\
           &+& \hat{\eta}(p) \hat{\eta}(-p) B(p) + \hat{\Phi}(p)
           \hat{\eta}(-p) C(p) \},
\label{29}
\eeqn

\noindent where

 \beq
     A(p) = \frac{g^2}{2\pi}~ p^2 + 
     \frac{1}{2}[\hat{b}_{(0)}(p) p_1^2 +
           \hat{b}_{(1)}(p) p_0^2],
\label{30}
\eeq

\beq
B(p) = \frac{1}{2}[\hat{b}_{(0)}(p) p_0^2 +
           \hat{b}_{(1)}(p) p_1^2],
\label{31}
\eeq

\beq
C(p) = [\hat{b}_{(0)}(p) - \hat{b}_{(1)}(p)] p_0 p_1,
\label{32}
\eeq
                                         
\noindent and $\hat{\Phi},\hat{\eta}$ and $\hat{b}_{(\mu)}$ are the Fourier
transforms of $\Phi, \eta$ and $b_{(\mu)}$ respectively.

\noindent Eq. (\ref{29}) can be easily diagonalized through the change 

\beqn
\hat{\Phi} & = & \hat{\zeta} - \frac{C}{2A}\hat{\xi} \nonumber\\
\hat{\eta} & = & \hat{\xi}.
\label{33}
\eeqn

\noindent We then have the following propagators for $\hat{\zeta}$ 
and $\hat{\xi}$:

\beqn
G^{-1}_{\zeta}(p) & = & \lambda p^2 + \frac{1}{2} [\hat{b}_
{(0)} p_1^2 + \hat{b}_{(1)} p_0^2]\\
G^{-1}_{\xi}(p) & = & \frac{\lambda p^2 [\hat{b}_{(0)} p_0^2 +
           \hat{b}_{(1)} p_1^2] + \frac{\hat{b}_{(0)}\hat{b}_{(1)}}{2} p^4} 
           {2 \lambda p^2 + \hat{b}_{(0)} p_{1}^2 + \hat{b}_{(1)} p_{0}^2},
\label{34}
\eeqn

\noindent where $\lambda = \frac{g^2}{2 \pi}$ and $\hat{b}_{(0)}$, 
$\hat{b}_{(1)}$  are functions of p.
These expressions are further simplified in the case 
$\hat{b}_{(0)}$ =$ \hat{b}_{(1)}$. In particular, when
$\hat{b}_{(0)}$ =$ \hat{b}_{(1)} \propto \frac{1}{p^2}$ the
$\hat{\zeta}$ field acquires a mass, whereas $\hat{\xi}$ 
becomes a non propagating field.
\newpage

\section{The Tomonaga-Luttinger model}

\hs  The approach depicted above will be now applied to the TL model  
\cite{Mattis} \cite{Luttinger} \cite{Tomonaga}. This model 
describes a non-relativistic gas of spinless particles 
(electrons) in which the free dispersion relation is taken to be linear. 
The free-particle Hamiltonian
is given by

\beq
H_0 = v_{F} \int dx \Psi^{\dagger}(x) (\sigma_{3} p - p_{F}) \Psi(x)
\label{48}
\eeq

\noindent where $v_{F}$ and $p_{F}$ are the Fermi velocity and momentum
respectively ($v_{F}p_{F}$ is a convenient origin for the energy scale). 
$\sigma_{3}$ is a Pauli matrix and $\Psi$ is a column bispinor with 
components $\Psi_1$ and $\Psi_2$
($\Psi^{\dagger} = (\Psi_1^{\dagger}~~\Psi_2^{\dagger})$). 
The function 
$\Psi_1(x) ~[\Psi_2(x)]$ is associated with the motion of particles in 
the positive [negative] $x$ direction. The interaction piece of
the Hamiltonian, when only forward scattering is considered, is

\beq
H_{int} = \int dx \int dy \sum_{a,b} \Psi^{\dagger}_a (x) \Psi_a(x)
V_{ab}(x,y) \Psi^{\dagger}_b(y) \Psi_{b}(y)
\label{49}
\eeq

\noindent where $a,b=1,2$, and the interaction matrix is parametrized
in the form

\beq
V_{ab} = \left( \begin{array}{cc}
            v_1 & v_2\\
            v_2 & v_1
            \end{array} \right).
\label{50}
\eeq 

\noindent Using the imaginary-time formalism one can show that 
the finite-temperature
\cite{Ma} \cite{BDJ} action for this problem becomes

\beqn
S_{TL}& =& \int^{\beta}_{0}d\tau \int dx~ \{\bP \gamma_0 (\partial_{\tau} -
v_p p_F) \Psi + v_F \bP \gamma_1 \partial_x \Psi\}\nonumber\\
& + &
\int^{\beta}_{0} d\tau \int dx \int dy 
\sum_{a,b} \Psi^{\dagger}_a \Psi_a(x,\tau) V_{ab}(x,y) \Psi^{\dagger}_b
\Psi_b(y,\tau).
\label{51}
\eeqn

For simplicity, in this Section we shall set $v_F=1$ and consider the case 
$v_1 = v_2$ in (\ref{50}) \cite{Mattis}. We shall also restrict ourselves 
to the zero temperature limit ($\beta \rightarrow \infty$). Under these 
conditions it is easy to verify that $S_{TL}$ coincides with the non-local
Thirring model discussed in the precedent Section, provided that the 
following identities hold:

\beqn 
g^2& =& 2\nonumber\\ 
V_{(0)}(x,y)& =& v_1(x,y) = v_2(x,y)= v( x_1 - y_1)
\delta (x_0 - y_0)\nonumber\\
V_{(1)}& =& 0
\label{52}
\eeqn

\noindent Of course one has also to make the shift 
$\bP \gamma_0 \partial_0 \Psi 
\rightarrow \bP \gamma_0 (\partial_0 - p_F ) \Psi$ and identify 
$x_0 = \tau $, $x_1 = x$.

One then can employ the method described in the precedent Section in order 
to study the Tomonaga-Luttinger model. This model has been previously studied, 
through a different functional approach, by
D.K. Lee and Y. Chen \cite{LC}. These authors, however, avoided the 
use of the decoupling technique presented here. Our approach 
is particularly useful when considering spin-flipping
interactions, i.e. the non-Abelian extension of the model. For simplicity I
will not consider this case in this work, but the interested reader will 
find related discussions in refs. \cite{1} and \cite{2}.

Let us first focus our attention to the dispersion relations corresponding
to the elementary excitations of the model at hand. These states correspond
to the normal modes whose dynamics is governed by the action (\ref{29}). As 
it is well-known, the spectrum of these modes is obtained from 
the poles of the 
corresponding propagators. Alternatively, one can write the effective 
Lagrangian as 

\beq 
L_{eff} = \frac{1}{(2 \pi)^{2}}\left ( \begin{array}{cc} 
\hat{\Phi}~ \hat{\eta} \end{array} \right)
\left( \begin{array}{cc}
A & C/2 \\
C/2 & B \end{array} \right) \left(\begin{array}{cc} \hat{\Phi} \\
\hat{\eta} \end{array} \right)
\label{53}
\eeq

\noindent (with A, B and C defined in (\ref{30})-(\ref{32})) and solve the 
equation 

\beq 
\Delta(p) = 0, \\
\label{54}
\eeq 

\noindent with $\Delta (p)= C^2(p) - 4 A(p) B(p)$.
Going back to real frecuencies : $p_{0} = i\omega $, $p_{1} = q $,
(\ref{54}) 
yields a biquadratic equation for $\omega$. The relevant solution is

\beq 
\omega_{-}^{2}(q) = \frac{\hat{b}_{(1)}}{\hat{b}_{(0)}} \frac{ 2 \lambda + 
\hat{b}_{(0)}}
{ 2 \lambda + \hat{b}_{(1)}} q^{2}. \\
\label{56}
\eeq

Inserting now the identities (\ref{52}) in ( \ref{56})
($\lambda =\frac{ g^{2}}{2 \pi}$) we obtain

\beq 
\omega_{-}^{2}(q) = q^{2} \{ 1 + \frac{2 v(q)}{ \pi} \} 
\label{57}
\eeq

\noindent which is the well-known result for the spectrum of the 
charge-density 
excitations of the TL model in the Mattis-Lieb version \cite{Mattis}.

The next step is to compute the electron propagator. To this end, having 
established the correspondence between the TL and NLT 
models, we can use the decoupling technique at the level of the 2-point
function. As usual, the non-vanishing components of the fermionic 2-point 
function are factorized into fermionic and bosonic contributions. Carefully 
taking into account the Fermi momentum in the free-fermion 
factor one gets

\beq 
G^{0}_{\pm}(z) = \frac{ e^{ \pm ip_{F}z_{1}} (z_{0} \pm i z_{1})}
{2 \pi \mid z \mid^2}
\label{58}
\eeq

\noindent whereas the bosonic factor becomes

\beq 
B_{\pm}(z) = exp[ \frac{1}{ \pi^{2}} \int d^2p~ \frac{v(p)}{p^2} 
\frac{ sen^2(\frac{p.z}{2})( p_{0} \pm i p_{1})^{2} }
{ p_{0}^{2} + ( 1 + 2v/\pi)p_{1}^{2} }].
\label{59}
\eeq

\noindent The momentum distribution for branch 1 (2) electrons 
is given by

\beq 
N_{ \stackrel{1}{2}}(p_1) = {\tt C(\Lambda)}
\int_{- \infty}^{ \infty} dz_{1}~ e^{-ip_{1}z_{1}}
\lim_{z_{0} \rightarrow 0} G_{ \pm}(z_{0}, z_{1})
\label{60}
\eeq

\noindent Replacing ( \ref{58}) and ( \ref{59}) in ( \ref{60}) we get

\beqn
N_{ \stackrel{1}{2}}(p_{1})& =& \pm \frac{{\tt C(\Lambda)} i}{2 \pi} 
\int_{- \infty}^{ \infty} 
dz_{1}~ e^{-i z_{1}( p_{1} \mp p_{F})} \times \nonumber \\
& \times & \frac{1}{z_{1}} 
exp\{ \frac{1}{ \pi^2}
\int \frac{d^{2}p~ v(p)}{p^{2}} sen^{2} (\frac{p_{1}z_{1}}{2}) 
\frac{(p_{0}^{2}
-p_{1}^{2})}{p_{0}^{2} + [ 1 + 2 v(p)/ \pi]p_{1}^{2}}\}.
\label{61}
\eeqn

\noindent Here ${\tt C(\Lambda)}$ is a normalization constant 
depending on an ultraviolet cutoff $\Lambda$.
In the local limit, in which $ v(p) = const$, the integrals 
in the momentum can be easily evaluated and one 
obtains

\beq 
N_{ \stackrel{1}{2}}(p_{1}) = \pm \frac{i}{2 \pi} {\tt C(\Lambda)} 
\int_{- \infty}
^{ \infty} dz_{1} \frac {e^{-i (p_{1} \mp  p_{F}) z_{1}}}{z_{1}^{1 + \sigma}}
\label{62}
\eeq

\noindent with 

\beq
\sigma = \frac{1}{2} \{(1 + \frac{2 v}{ \pi})^{1/2} + 
( 1 + \frac{2 v}{ \pi})^{-1/2} - 2 \}
\label{63}
\eeq

\noindent Note that in the free case $ v \rightarrow 0$ one gets 
$ \sigma = 0$, which
leads to the well-known normal Fermi-liquid behavior, 

\beq
N_{ \stackrel{1}{2}} \propto \theta (p_{1} \pm p_{F}).
\label{64}
\eeq 

As soon as the interaction is switched on, one has $ \sigma \neq 0$ and the 
Fermi edge singularity is washed out, giving rise to the so called 
Luttinger-liquid behavior \cite{Haldane}. It has been emphasized 
recently \cite{Hu} that the experimental data obtained for
one-dimensional structures can be succesfully explained on the basis
of standard Fermi-liquid theory. We believe that our 
approach could be useful to explore some modifications of the TL model 
to take into account, for instance, the presence of impurities or defects, 
that might yield a restoration of the edge singularity. The issue of how to
incorporate impurities in our framework will be briefly addressed in the 
next Section.

\newpage
 
\section{Fermionic impurities}

The main purpose of this Section is to comment on the extension of the path-integral 
approach to non-local bosonization proposed in \cite{1}, to the case in
which an interaction between the electrons and a finite density of fermionic
impurities is included in the action. This generalization of the non-local
bosonization procedure provides a new way to examine the low-energy physics
of the TL model in the presence of localized impurities, that could allow to
make contact with recent very interesting studies \cite{Sch'},\cite{GiS} on
the response of a Luttinger liquid to localized perturbations.

We describe the impurities following the work of Andrei \cite{Andrei}, 
who introduced a new fermionic field with vanishing kinetic energy to represent 
a finite density of impurities, arbitrarily (not randomly) 
situated. This treatment has been previously employed, for example, in the 
path-integral bosonization of the Kondo problem \cite{FS}.

We introduce a non-local diagonal potential matrix binding impurities 
and electrons through their corresponding fermionic currents. This procedure 
allows to treat a wide range of possible interactions, depending on 
the precise functional form of the potential matrices. 
The complete coupling term includes interactions between charge, current, 
spin and spin-current densities. Our functional approach enables us to obtain 
an effective action governing the dynamics of the collective modes, providing 
then a practical framework to face a non-perturbative analysis of bosonic 
degrees of freedom in the presence of impurities.

We start from the partition function
\begin{equation}
Z = \int D\bar{\Psi}~D\Psi~D\bar d~Dd~e^{-S},
\label{a}
\end{equation}
where the action $S$ can be splitted as
\begin{equation}
S = S_0 + S_{int},
\end{equation}
with
\begin{equation}
S_0  =  \int d^2x~ [\bar{\Psi} i \raise.15ex\hbox
{$/$}\kern-.57em\hbox{$\partial$} \Psi + d^{\dagger} i \partial_t d] 
\end{equation}
and
\begin{equation}
S_{int} = - \int d^2x~d^2y~ [J^a_{\mu}(x) V^{ab}_{(\mu)}
(x,y)J^b_{\mu}(y) + J^a_{\mu}(x) U^{ab}_{(\mu)}(x,y) S^b_{\mu}
(y)],
\label{b}
\end{equation}
where the electron field $\Psi$ is written as
\[ \Psi = \left( \begin{array}{c} 
              \Psi_{1} \\
              \Psi_{2} \\
              \end{array} \right),  \]
with $\Psi_1$ ($\Psi_2$) in the fundamental representation of  
U(N), describing right (left) movers, whereas the impurity field $d$
is given by
\[ d = \left( \begin{array}{c}  
                            d_{1}  \\
                            d_{2}  \\
                            \end{array} \right). \]
Note the absence of a spatial derivative in the free piece of the 
impurity action, meaning that the corresponding kinetic energy is zero.
Concerning the electronic kinetic energy, we have set the Fermi velocity
equal to 1.\\
The interaction pieces of the action have been written in terms of 
U(N) currents $J_{\mu}^{a}$ and $S_{\mu}^{a}$, defined as
\begin{eqnarray}
J_{\mu}^{a} &=& \bar{\Psi} \gamma_{\mu} \lambda^{a} \Psi,  \nonumber\\
S_{\mu}^{a} &=& \bar d \gamma_{\mu} \lambda^{a} d,~~~ a = 0,1,...,N^2-1,
\label{,}
\end{eqnarray}
with $\lambda^0= I/2$, $\lambda^j= t^j$, $t^j$  being the SU(N)
generators normalized according to $tr (t^i t^j ) = \delta^{ij}/2$. 
$V^{ab}_{(\mu)}(x,y)$ and $U^{ab}_{(\mu)}(x,y)$ are $N^2 \times 
N^2$ matrices whose elements are symmetric bilocal arbitrary potentials 
describing the electron-electron (e-e) and the electron-impurity (e-i) 
interactions, respectively.  

Although we have obtained a bosonized effective action for the general
(non-abelian, spin-flipping) problem (See \cite{2} for details), here I shall
restrict myself to the maximal abelian subgroup of U(2). In this case the model 
describes a many-body system of spin-$\frac{1}{2}$ fermions when 
spin-flipping processes are not allowed. 
Now, the potential matrices are diagonal whose elements can be written in 
terms of the g-functions defined by S\'olyom \cite{Solyom} as
\begin{eqnarray}
V_{(0)}^{00}&=&\frac{1}{4}(g_{4 \parallel}+ g_{4 \perp} + g_{2 \parallel} 
+ g_{2 \perp}),\nonumber\\
V_{(0)}^{11}&=&\frac{1}{4}(g_{4 \parallel}- g_{4 \perp} + g_{2 \parallel} 
- g_{2 \perp}),\nonumber\\
V_{(1)}^{00}&=&\frac{1}{4}(-g_{4 \parallel}- g_{4 \perp} + g_{2 \parallel} 
+ g_{2 \perp}),\nonumber\\
V_{(1)}^{11}&=&\frac{1}{4}(-g_{4 \parallel} + g_{4 \perp} + g_{2 \parallel} 
- g_{2 \perp}).
\end{eqnarray}
It is straightforward to verify that the e-e interaction term in
(\ref{b}) contains the whole set of diagrams associated to forward scattering 
processes without spin-flips. Let us recall that the coupling constants for 
incident fermions with parallel spins are denoted by the susbscript $\parallel$ 
and that for fermions with opposite spins by the subscript $\perp$. In the $g_2$ 
processes the two branches (left and right moving particles) are coupled, while 
in the $g_4$ processes all four participating fermions belong to the same 
branch.
\noindent The Tomonaga-Luttinger model, with charge-density fluctuations only,
corresponds to $V_{(0)}^{11} = V_{(1)}^{11} = 0 $. In a completely analogous 
way we introduce the potentials that couple electron and impurity currents 
in the form
\begin{eqnarray*}
U_{(0)}^{0 0} & = & \frac{1}{4} ( h_{4\parallel} + h_{4\perp} + h_{2\parallel} 
+ h_{2\perp}), \\
U_{(0)}^{11} & = & \frac{1}{4} ( h_{4\parallel} - h_{4\perp} + h_{2\parallel} 
- h_{2\perp}) , \\
U_{(1)}^{0 0} &  = & \frac{1}{4} (- h_{4\parallel} - h_{4\perp} + h_{2\parallel} 
+ h_{2\perp}),  \\
U_{(1)}^{11} & = & \frac{1}{4} (- h_{4\parallel} + h_{4\perp} + h_{2\parallel} 
- h_{2\perp} ).
\end{eqnarray*}
This description includes both charge and spin density interactions, as well as 
spin-current interactions. A Kondo-like interaction, i.e. the coupling between 
spin densities only, corresponds to the case $U_{(0)}^{00} = U_{(1)}^{00} = 0$.

\indent In order to carry out the bosonization of the model one has to face some
technical difficulties that we shall not describe here (Again, see \cite{2}
for details). Let me say that, in this case one has two fermionic determinants in 
the partition function, one associated to electrons and a new one, related to
the impurity degrees of freedom. When they are conveniently decoupled, an
effective action for the collective modes of the system is again obtained.
In other words, one is left with
a partition function in terms of  
bosonic fields $\Phi^i$ and $\eta^i$ which, by comparison with the 
impurity-free case, one naturally identifies with the collective modes of
the system. The result is
\begin{equation}
 Z = \int D\Phi^i~ D\eta^i~exp - \left\{ S_{eff}^{00} + S_{eff}^{11} \right\}, 
 \label{87}
 \end{equation}
 where the actions, in Fourier space, are given by
\newpage
 \begin{eqnarray}
 S_{eff}^{ii} & = & \frac{1}{(2 \pi)^{2}} \int d^2p 
[ \hat{\Phi}^{i}(p) A^{ii}(p) \hat{\Phi}^{i}(-p) + 
\hat{\eta}^{i}(p) B^{ii}(p) \hat{\eta}^{i}(-p)  + \nonumber \\
& + & \hat{\Phi}^{i}(p) \frac{C^{ii}(p)}{2} \hat{\eta}^{i}(-p) + 
\hat{\eta}^{i}(p) \frac{C^{ii}(p)}{2} \hat{\Phi}^{i}(-p) ], 
\label{88} 
\end{eqnarray}
where 
\begin{eqnarray}
 A(p) & = & \frac{1}{\Delta(p)}\left\{ \frac{p^{2}}{\pi} \Delta - a_{0} a_{1} 
 \frac{p_{1}^{2}}{\pi} + \frac{1}{2 \pi} (a_{0}^{2} p_{1}^{2} - a_{1}^{2} 
 p_{0}^{2}) -2 a_{0}^{2} a_{1}^{2} (\frac{p_{1}^{2}}{ b_{1}} + \frac{p_{0}^{2}}
 {b_{0}}) \right\}, \nonumber \\   
B(p) & = & \frac{1}{\Delta(p)}\left\{ \frac{p_{1}^{2}}{\pi} a_{0} a_{1} + 
\frac{1}{2 \pi} (a_{0}^{2} p_{0}^{2} - a_{1}^{2} p_{1}^{2}) - 2 a_{0}^{2} 
a_{1}^{2} ( \frac{p_{1}^{2}}{b_{0}} + \frac{p_{0}^{2}}{b_{1}}) \right\}, 
\nonumber \\
C(p) & = & \frac{1}{\Delta(p)}\left\{ \frac{ a_{0} a_{1}}{\pi} ( \frac{p_{1}^
{3}}{p_{0}} - p_{0} p_{1} ) + \frac{ p_{0} p_{1}}{\pi} (a_{0}^{2} + a_{1}^{2}) + 
4 p_{0} p_{1} a_{0}^{2} a_{1}^{2} (\frac{1}{ b_{0}} - \frac{1}{b_{1}})\right\},
\nonumber \\
\Delta (p)& =& \frac{ p_{1}^{2}}{ 4 \pi^{2} p_{0}^{2}} + 4 ( \frac{1}{ 4 \pi} -
\frac{ a_{1}^{2}}{ b_{1}}) ( \frac{1}{ 4 \pi } + \frac{a_{0}^{2}}{ b_{0}}).
\label{88a}
\end{eqnarray}
\hspace*{0.6cm} For the sake of clarity we have omitted $ii$ superindices in 
the above expressions, which are written in terms of the Fourier transforms 
of the inverse potentials. (Note that 
$b_{\mu}(p) = V_{(\mu)}^{-1}(p)$ and $a_{\mu}(p) = U_{(\mu)}^{-1}(p)$ ).

\indent This is one of our main results. We have obtained a completely bosonized 
action for the collective modes corresponding to a system of electrons which 
interact not only between themselves, but also with fermionic localized
impurities at $T=0$.
This effective action describes the dynamics of charge density 
($\Phi^0$ and $\eta^0$) and spin density ($\Phi^1$ and $\eta^1$)fields. 
As we can see, these modes remain decoupled as in the impurity free case. 
Their dispersion relations can be obtained from the poles of the corresponding 
propagators. Alternatively, one can write the effective Lagrangian as
\begin{equation}
L_{eff}^{ii} = \frac{1}{2\pi} \left( \hat{\Phi}^i ~ \hat{\eta}^i \right)
\left( \begin{array}{cc}
A^{ii} & C^{ii}/2 \\
C^{ii}/2& B^{ii}
\end{array} \right) ~ \left( \begin{array}{c}
\hat{\Phi}^i\\
\hat{\eta}^i
\end{array}\right),
\end{equation}
with $A,B$ and $C$ as defined above, and solve the equation 
\begin{equation}
C^2(p) - 4 A(p) B(p) = 0.
\end{equation}
Going to real frecuencies: $p_0 = i \omega, p_1 = q$, this equation
has the following pair of relevant solutions:
\begin{equation}
\omega^{2}_{\rho}(q) = q^{2}  \frac{ 1 + \frac{2}{\pi} V_{(0)}^{00} + 
\frac{1}{2 \pi ^{2}}\{(U_{(0)}^{00})^{2} - 2 U_{(0)}^{00} U_{(1)}^{00}\} }
{1 + \frac{2}{\pi} V_{(1)}^{00} - \frac{1}{2 \pi ^{2}} 
(U_{(1)}^{00})^{2}},
 \label{141}
 \end{equation}
\begin{equation}
\omega^{2}_{\sigma}(q) = q^{2}  \frac{ 1 + \frac{2}{\pi} V_{(0)}^{11} + 
\frac{1}{2 \pi ^{2}}\{(U_{(0)}^{11})^{2} - 2 U_{(0)}^{11} U_{(1)}^{11}\} }
{1 + \frac{2}{\pi} V_{(1)}^{11} - \frac{1}{2 \pi ^{2}} 
(U_{(1)}^{11})^{2}},
\label{15'}
\end{equation}
The first equation gives the dispersion relation associated to 
charge-density fluctuations $(\hat{\Phi}^0,\hat{\eta}^0)$, whereas 
the second one corresponds to spin-density modes $(\hat{\Phi}^1,
\hat{\eta}^1)$. 

As a confirmation of the validity of our approach, we note that the above 
dispersion relations, involving both e-e and e-i interaction potentials, 
coincide with the well-known result for the spectrum of charge and spin 
excitations in the TL model without impurities, obtained by choosing 
$V_{(1)} = U_{(0)} = U_{(1)} =0$ and $V_{(0)} = v(q)$, in the above 
formulae.

\indent Let us now consider the fermionic 2-point function
\begin{equation}
\langle \Psi (x) \bar{\Psi} (y) \rangle = \left( \begin{array}{cc}
                                           o & G_1(x,y)\nonumber\\
                                    G_2(x,y) & 0 
                                          \end{array}  \right)
\end{equation}
where 
\begin{equation}
 G_{1(2)}(x,y) = \left( \begin{array}{cc}
                        G_{1(2)\uparrow}(x,y) & 0\nonumber\\
                                            0 & G_{1(2)\downarrow}(x,y) 
                        \end{array} \right)
\end{equation}
The subindex $1 (2)$ means that we consider electrons belonging to 
the branch $1 (2)$, and $\uparrow$ ($\downarrow$) indicates that the field 
operator carries a spin up (down) quantum number.
Let us recall that in the present case we have disregarded those processes 
with spin-flip. This is why the fermionic Green function do not have non-zero
components with mixed spin indices.\\
To be specific we consider $G_{1 \uparrow}$ (similar 
expressions are obtained for $G_{2 \uparrow}$, $G_{1 \downarrow}$ and $G_{2 
\downarrow}$ ). When the decoupling chiral change is performed, the components 
of the Green functions are factorized into fermionic and bosonic contributions 
in the form
  \begin{eqnarray}
  G_{1 \uparrow}(x,y) & = & <\Psi_{1 \uparrow} (x) \Psi^{\dag}_{1 \uparrow}(y)>
\nonumber \\
  & = & G_{1 \uparrow}^{(0)} (x,y) <e^{\left\{[ \Phi^{0}(y) 
  -\Phi^{0}(x)] + 
i[\eta^{0}(y) - \eta^{0}(x)]\right\}}>_{00}  \times  \nonumber \\ 
& \times & <e^{\left\{[ \Phi^{1}(y) -\Phi^{1}(x)] + 
i[\eta^{1}(y) - \eta^{1}(x)]\right\}}  >_{11},
\label{101}
\end{eqnarray}
where $G_{1 \uparrow}^{(0)} (x,y)$ is the free propagator, which
involves the Fermi momentum $p_F$, and is given by
\begin{equation}
G_{1 \uparrow}^{(0)} (x,y) = \frac{e^{i p_F z_1}}{2 \pi \mid z \mid^2} 
(z_0 + i z_1).
\label{102}
\end{equation}
The symbol $ <>_{ii}$ means v.e.v. with respect to the action 
(\ref{88}, \ref{88a}). Exactly as we did in the impurity-free case 
(\cite{1}), the bosonic factors in (\ref{101}) can be evaluated by
appropriately shifting the fields. Indeed, working in momentum-space, and 
defining the non-local operator
\begin{equation}
D(p;x,y) = e^{-ip.x} - e^{-ip.y},
\end{equation}
the functional integrations can be performed, yielding 
\begin{eqnarray}
<\Psi_{1 \uparrow} (x) \Psi^{\dag}_{1 \uparrow}(y)> = G_{1 \uparrow}^{(0)} 
(x,y)&& exp\{-\int \frac{d^2p}{(2 \pi)^2} D^2 \frac{A^{00} - B^{00} + i 
C^{00}}{4 A^{00}B^{00} - (C^{00})^2}\}\nonumber\\
& & exp\{-\int \frac{d^2p}{(2 \pi)^2} D^2 \frac{A^{11} - B^{11} + i C^{11}}
{4 A^{11}B^{11} - (C^{11})^2}\},\nonumber\\
\end{eqnarray}
with $A^{ii}, B^{ii}$ and $C^{ii}$ given by (\ref{88a}).
In order to continue the calculation one needs, of course, to specify the
couplings and perform the integrals. This means that our formula could be used 
to test the effect of different e-e and e-i potentials on the behavior of the 
fermionic propagator.

\indent As a final illustration of our procedure, now I will 
show how to compute the electronic momentum distribution, for a quite peculiar
choice of e-e and e-i potential matrix elements.\\
Let us consider the momentum distribution of electrons belonging to branch 1 
and with spin-up. This distribution is given by
\begin{equation}
N_{1 \uparrow}(q) = C(\Lambda) \int^{+\infty}_{-\infty} dz_1~ e^{-iqz_1}
\lim_{z_0 \rightarrow 0} G_{1\uparrow}(z_0, z_1).
\label{103}
\end{equation}
We shall set
\[
V_{(1)}^{00} = V_{(1)}^{11} = 0,~\\
V_{(0)}^{00} = \frac{\pi}{2} r,~\\
V_{(0)}^{11} = \frac{\pi}{2} s,
\]
which corresponds to an e-e interaction including only 
charge-density fluctuations (the usual TL model). 
Concerning the interaction between electrons and impurities, 
we shall take into account only spin-density and spin-current interactions,
\[
U_{(0)}^{00} = U_{(1)}^{00} = 0,\\
\left( U_{(0)}^{11}\right) ^2 = \left( U_{(1)}^{11}\right) ^2 = 2 \pi^2 t.
\]
Note that for repulsive electron-electron interactions one has 
$r > 0$ and $s > 0$, whereas $t > 0$ for both ferromagnetic and 
antiferromagnetic couplings.\\
Taking the limit $z_0 \rightarrow 0$ in (\ref{101}) and replacing  eqs. 
(\ref{101}) and (\ref{102}) in (\ref{103}) one gets
\begin{equation}
 N_{1 \uparrow}(q) = C(\Lambda) \int dz_{1} \frac{e^{-i(q - p_{F})z_{1}}}{z_{1}}
 e^{-\int dp_{1} \frac{ 1 - cos p_1 z_1 }{p_1}  \Gamma (p_1)},
 \label{104}
 \end{equation}
where $C(\Lambda)$ is a normalization constant depending on an ultraviolet 
cutoff $\Lambda$, and $\Gamma(p_1)$ depends on $p_1$ through the potentials 
in the form
\begin{eqnarray}
\Gamma (r,s,t) & = & \frac{( \mid 1 + r \mid^{1/2} -  1)^{2}}
{\mid 1 + r \mid^{1/2}} + \frac{( \mid 1 + s - t \mid^{1/2} - \mid 1 -t 
\mid^{1/2})^{2}}{\mid 1 + s - t \mid^{1/2} \mid 1 - t \mid^{1/2}} -\nonumber\\
 & - & \frac{ 2 t ( t - t_{c} )}{  \mid 1 + s - t \mid^{3/2} 
\mid 1 - t \mid^{1/2}}.
\label{105}
\end{eqnarray}
If we define
\begin{equation}
f(s,t) = \frac{( \mid 1 + s - t \mid^{1/2} - \mid 1 -t \mid^{1/2})^{2}}
{\mid 1 + s - t \mid^{1/2} \mid 1 - t \mid^{1/2}}, 
\label{106}
\end{equation}
\begin{equation}
h(s,t)= t \frac{ 2 t ( t -1 ) + s (s - 1)}{  \mid 1 + s - t \mid^{3/2} 
\mid 1 - t \mid^{1/2}},
\label{107}
\end{equation}
we can write eq. (\ref{105}) in the form
\begin{equation}
\Gamma (r,s,t) = f(r,0) + f(s, t) - h(s,t).
\label{108}
\end{equation}
Once again, in order to go further and make the integration in
(\ref{104}), one has to specify the functional form of $r$, $s$ and $t$. 
At this point one observes that we are in a position of discussing, through 
this simple example, an interesting aspect of the general model under 
consideration. Indeed, we can try to determine under which conditions it is 
possible to have a restoration of the Fermi edge. To this end, and as a first 
approximation, we shall consider contact interactions ($r$, $s$ and $t$ 
constants) and search for those relations between potentials giving 
$\Gamma (r,s,t) = 0$. 
In this last case one obtains the well-known normal 
Fermi-liquid (FL) behavior
\begin{equation}
N_{1 \uparrow}(q) \approx \Theta (q - p_F).
\label{109} 
\end{equation}
At this point some remarks are in order. 
In the impurity free case ($t = 0$), $\Gamma (r,s,t)$ cannot vanish for any 
value of $r$ and  $s$ other than $r=s=0$, which corresponds to the 
non-interacting Fermi gas. This result is consistent with the well-known
LL behavior of the TL model.\\
In order to have collective modes with real frecuencies ($\omega^2>0$), one 
finds two regions where the FL edge could be restored: $t > 1+s$ and $t<1$.

\noindent In eq.(\ref{106}) one can observe that $f(r,0)>0$, thus setting 
$\Gamma = 0$ yields the condition
\begin{equation}
F(s,t) = h(s,t) - f(s,t) > 0. 
\label{110}
\end{equation}
A simple numerical analysis of $F(s,t)$ shows that the above 
inequality is not fulfilled for $0 < t <1$. The electron-impurity coupling is 
not strong enough in this region as to eliminate the LL behavior. On the 
contrary, for $t > 1 + s$ equation (\ref{106}) can be always satisfied. 
Moreover, in this region, we obtain a surface in which the condition 
$\Gamma = 0$ provides the following analytical solution for $r$ in terms of 
$F(s,t)$
\begin{equation}
r = F^{2}/2 + 2 F + ( 1 + F/2) ( F^{2} + 4 F)^{1/2}.
\label{111}
\end{equation}
The above discussion can be summarized by identifying the following
three regions in the space of couplings:

\noindent Region I, given by $0 \leq t < 1$, in which one necesarilly has LL 
behavior. Region II, with $ 1 \leq t < s + 1$, in which the frecuency of the 
spin density excitations becomes imaginary; and region III, given by 
$t > s + 1$, where the FL behavior is admitted. In this region Eq(\ref{107}) 
defines a surface in the space of potentials on which FL behavior takes place. 
One particular solution belonging to this surface is obtained by choosing 
$s = 0$ in (\ref{111}), which yields 
\[ r(t) = \frac{2t(3t-2) + 2(2t-1)\sqrt{3t^2 - 2t}}{(t-1)^2}.\]
corresponding to the case in which the dispersion relation of the 
spin density excitations is given by $\omega^2 = q^2$. For $t$ large, $r$ 
approaches a minimum value $r_{min}= 6 + 4\sqrt{3}$, a feature that is shared 
with each curve $ s = constant$ on the "FL surface". \\
\indent In summary, we have shown how the e-i couplings can be tuned in order to 
have a restoration of the Fermi edge in a TL model of electrons interacting 
with fermionic impurities. Unfortunately, we could analytically work out this
mechanism only for a very peculiar choice of the e-i couplings, which 
evidently weakens its experimental relevancy. Besides, a more realistic study 
should include at least the backward-scattering processes. However, we think 
this discussion deserves attention as a first step towards a possible 
reconciliation between the standard TL model and the FL phenomenology.
\newpage
\section{Comments on further developments}

I would like to end by making some remarks on recent works that
help to improve our present understanding of the Thirring-like model 
discussed in this talk.

In ref.\cite{3} we addressed our attention to the vacuum properties of the
above mentioned model. As it is well-known, ground-state wave
functionals (GSWF's) have in general very complex structures. Due to
this fact, their universal behavior has been seldom explored in the
past. Fortunately, in a recent series of papers, an alternative way to
compute GSWF's was presented \cite{LF} \cite{F} \cite{FMS}. By
conveniently combining the operational and functional approaches to
quantum field theories, these authors provided a systematic
path-integral method that, at least in the context of $1+1$ systems,
seems to be more practical than the previously known semiclassical
and Bethe ansatz techniques. We take advantage of these advances and applied
them to shed some light on the vacuum structure of the NLT. In particular 
we got a closed formula that
gives the probability of the vacuum state as a functional, not only of
the density configuration but also of the potentials that bind the
original fermionic particles of the system. This result allowed us to
find a non-trivial symmetry of this vacuum with respect to the
interchange of density-density and current-current potentials.
Of course, this symmetry does not persist at the level 
of the dispersion relations of the collective
modes (plasmons), to which the excited states are expected to contribute.
We have also computed the general electromagnetic response of the model and 
the asymptotic behavior of GSWF's and density-waves frequencies for a wide 
variety of power-law potentials. This allowed us to identify different phases contained
in the non-local Thirring model.

In ref.\cite{4} we studied the non-local
Thirring model with a relativistic fermion mass term included in the 
action. Performing a perturbative expansion in the mass parameter,
we found that the NLT is equivalent to a purely bosonic action which is a
simple non-local extension of the sine-Gordon model. Thus, we have 
generalized Coleman's equivalence \cite{Coleman} to the case in which
the usual Thirring interaction is point-splitted through bilocal potentials.
In the language of many-body, non-relativistic systems, the relativistic mass
term can be shown to represent not an actual mass, but the introduction of
backward-scattering effects \cite{Back}. Therefore, our result could provide 
an alternative route to explore the non-trivial dynamics of (gapped) 
collective modes.

Finally, using the results of ref.\cite{5} we have recently examined the 
finite-temperature extension of the NLT, which allowed us to discuss 
termodynamical and transport properties of the 1d electronic system \cite{6}.

\newpage
\vspace{1 cm}

\underline{Acknowledgements}:  C.M.N. is partially supported
by CONICET, Argentina.

\end{document}